\documentclass[aps,prb,twocolumn,showpacs,superscriptaddress,
preprintnumbers,amsmath,amssymb]{revtex4}

\usepackage{graphicx}% Include figure files
\usepackage{dcolumn}% Align table columns on decimal point
\usepackage{bm}% bold math
\usepackage{ifthen}
\usepackage{booktabs}
\usepackage{SIunits}
\usepackage{ulem}

\usepackage{tabularx}
\usepackage{amsmath}
\usepackage{amssymb}
\usepackage{graphicx}
\usepackage{dcolumn}
\usepackage{bm}
\usepackage{color}

\begin{document}
\title{Temperature Dependent Empirical Pseudopotential Theory For Self-Assembled Quantum Dots}
\author{Jianping Wang}
\affiliation{Key Laboratory of Quantum Information, University of Science and Technology of
China, Hefei, 230026, People's Republic of China}
\author{Ming Gong}
\email{gong.skylark@gmail.com.}
%\affiliation{Key Laboratory of Quantum Information, University of Science and Technology of
%China, Hefei, 230026, People's Republic of China}
\affiliation{Department of Physics and Astronomy, Washington State University, Pullman,
Washington, 99164 USA}
\author{Guang-Can Guo}
\affiliation{Key Laboratory of Quantum Information, University of Science and Technology of
China, Hefei, 230026, People's Republic of China}
\author{Lixin He}
\email{helx@ustc.edu.cn.}
\affiliation{Key Laboratory of Quantum Information, University of Science and Technology of
China, Hefei, 230026, People's Republic of China}
\date{\today}

\begin{abstract}
We develop a temperature dependent empirical pseudopotential theory to study the
electronic and optical properties of self-assembled quantum dots (QDs) at finite
temperature. The theory takes the effects of both lattice expansion and lattice
vibration into account. We apply the theory to the InAs/GaAs QDs.  For the
unstrained InAs/GaAs heterostructure, the conduction band offset increases
whereas the valence band offset decreases with increasing of the temperature,
and there is a type-I to type-II transition at approximately 135 K.  Yet, for
InAs/GaAs QDs, the holes are still localized in the QDs even at room
temperature, because the large
lattice mismatch between InAs and GaAs greatly enhances the valence band
offset. The single particle energy levels in the QDs show strong temperature
dependence due to the change of confinement potentials.  Because of the changes
of the band offsets, the electron wave functions confined in QDs increase by
about 1 - 5\%, whereas the hole wave functions decrease by about 30 - 40\% when
the temperature increases from 0 to 300 K.  The calculated recombination energies
of exciton, biexciton and charged excitons show red shifts with increasing of
the temperature, which are in excellent agreement with available experimental
data.
\end{abstract}
\maketitle

\section{Introduction}
\label{sec-introduction}

During the past two decades, enormous progress has been achieved
in understanding the electronic and optical
properties of self-assembled quantum dots (QDs) both through theory and
experiments,
stimulated by their potential applications in QDs laser at room
temperature, \cite{Esaki73, Arakawa82, Kirstaedter94} and as qubits and
quantum photon emitters at low temperature. \cite{Imamoglu94, Lounis05,
Benson00, Stevenson06, Akopian06} For the former applications, it generally
requires high density and highly uniform QDs. The QD laser has been demonstrated
with much lower threshold current $J_c$ and much higher material and
differential gains as compared to the semiconductor quantum well lasers.
\cite{ledentsov00} For the latter applications, the preparation of single QD is
crucial. A number of methods have demonstrated the feasibility to isolate
single QD from QDs ensemble. \cite{bayer, marzin, yamauchi, tang} Rabi
oscillation of exciton and charged exciton \cite{rabi, CPT} in single QD have
been observed experimentally, showing that the charge and spin quantum states 
in single QD can be coherently controlled via optical method. The single
and entangled photon emission from single QD have also been
demonstrated experimentally, \cite{Shields-review} that are much brighter
\cite{ultrabright} than the traditional parameter-down entangled
photon source. \cite{Shih88, Kiess93} These experimental achievements pave the
way for future application of QDs in quantum computation.

On the other hand, the development of the atomistic theories, including the
empirical pseudopotential method \cite{zunger01, wang99a, wang99b} and the
tight-binding models \cite{zielinski10,lee01,santoprete03} provide deep
insight to the electronic and optical properties of self-assembled QDs.
The atomistic theories of QDs not only give results that agree well with
experiments, \cite{Ding, Singh09, Ediger07a, Ediger07b} but also greatly improve
our understanding of the properties of QDs.  The atomistic models capture
the correct point group symmetry of the QDs, which is missing in the continuum
model.  Therefore they can give correct interpretation of some subtle properties
of the QDs, e.g., the fine structure splitting (FSS), \cite{bester03} and light
polarization of excition. \cite{gong11} Unfortunately, so far all the theories of
QDs have been restricted to zero temperature.

Temperature is a very important degree of freedom in experiments to tune the
electronic and optical properties of QDs. For example, in QDs laser, the
temperature has be used to tune the laser wavelength.
\cite{Esaki73,Arakawa82,Kirstaedter94} In the QD-cavity system, the temperature
is generally used to tune the resonance between the QDs and the cavity in order
to achieve strong coupling between the two quantum systems. \cite{loo,
press, laussy, Englund} The temperature dependent optical spectra of (single and
ensemble) QDs have been investigated intensively in experiments in the
past decades. \cite{ortner, polaron1, ortner, bayer, marzin, yamauchi, rabi,
CPT, Yeo11} New physics, for example, the formation of excitonic
polaron\cite{polaron1, polaron2, polaron3}  which is due to  strong coupling
between exciton and optical phonons, may be found in QDs at high temperature.
However, a theoretical understanding of the temperature effects in QDs is still
missing. Therefore, to facilitate the future device applications of QDs, the
development of a temperature dependent theory is not only of theoretical
interest, but also of practical importance. 

In this work, we develop such a temperature dependent atomistic pseudopotential
theory to study the electronic and optical properties of QDs at finite
temperature.  We take the effects of both lattice expansion and lattice
vibration into account. The latter is done by introducing a temperature
dependent dynamical Debye-Waller factor to the structure factor. We first
examine the temperature dependent electronic structures of bulk InAs and GaAs,
and then apply the theory to investigate the electronic and optical properties
of self-assembled InAs/GaAs QDs. The calculated temperature dependent
photoluminescence (PL) spectra of QDs are in excellent agreement with available
experimental data. 

The rest of the paper is organized as follows. In Sec. \ref{sec-methods} we
introduce the temperature dependent empirical pseudopotential method (TDEPM).
In Sec. \ref{sec-bulk}, we study the electronic structures of bulk InAs, GaAs
using TDEPM, including the energy band gaps and band offsets, etc.  We present
the temperature dependent band offsets for InAs/GaAs QDs in Sec.
\ref{subsec-band offset} and the single particle energy levels and wave
functions of InAs/GaAs QDs in Sec. \ref{subsec-single}.  We discuss the
temperature dependent optical spectra of InAs/GaAs QDs in Sec. \ref{subsec-pl},
and summarize in Sec. \ref{sec-summary}.

\section{Methodology}
\label{sec-methods}

We consider InGaAs QDs embedded in the center of a 60
$\times$ 60 $\times$ 60 GaAs 8-atom unit cell. Periodic boundary condition is
used to obtain the single particle energy levels. 
To study the electronic and optical properties of the QDs at finite temperature,
we introduce temperature dependent pseudopotentials 
in the single-particle Hamiltonian,
\begin{equation}
\hat{H}=-\frac{1}{2}\nabla^{2}
+\sum_{n\alpha}\hat{v}_{\alpha}(\mathbf{r}-\mathbf{R}_{n\alpha}, \epsilon, T),
\label{eq-single-H}
\end{equation}
where $\hat{v}_{\alpha}(\mathbf{r}, \epsilon, T)$ is the strain dependent
screened empirical pseudopotential for atom of type $\alpha$ and atom
index $n$ at temperature $T$. 
${\mathbf{R}_{n\alpha}}$ is the optimized atom position from valance force field
(VFF) method. \cite{keating66, martin70} In a strained lattice, the atomic
potential is assume to have the form of,
\begin{equation}\label{eq:vr}
v_{\alpha}({\bf r},\epsilon, T) = v_{\alpha}({\bf r}, T)
[1 + \gamma_{\alpha} \text{Tr}(\epsilon({\bf r}))] \, ,
\end{equation}
where $\text{Tr}(\epsilon({\bf r}))$ is the local hydrostatic strain at ${\bf
r}$. $\gamma_{\alpha}$ is fitted to the deformation potentials of the bulk
materials.  The atomistic theory naturally captures the correct point group
symmetry of the QDs even at high temperature.

\begin{figure}[htbp]
\centering
\includegraphics[width=2.8in]{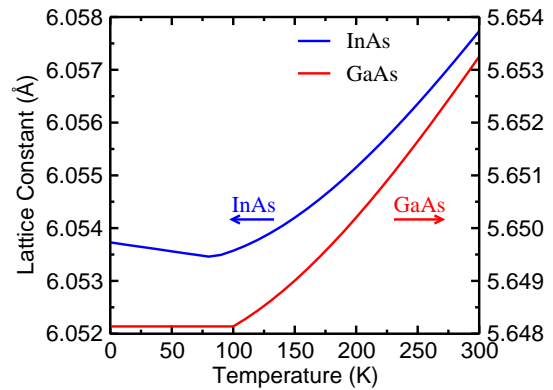}
\caption{(color online) Experimental temperature dependent lattice constants for InAs and GaAs.
The data are taken from Ref. \onlinecite{madelung2003}.}
\label{fig-lattice}
\end{figure}

The effects of lattice vibration can be taken into account using Brooks-Yu
theory, \cite{Yu-Brooks} which has been applied to study the energy gap of bulk materials
in the context of pseudopotentials. 
\cite{early-works-1, early-works-2, early-works-3, early-works-4} 
The total atomic potential at temperature $T$, 
which is the sum of all atomic potentials
$\hat{v}_{\alpha}(\mathbf{r}, \epsilon, T)$,  can be written as,
\begin{eqnarray}
V(\mathbf{r}, T) 
&& =\langle \sum_{\mathbf{q}} \sum_{\alpha} v_{\alpha}(\mathbf{q}) S({\bf q})
	e^{i\mathbf{q}\cdot\mathbf{r}}\rangle_T \nonumber \\
&& = \sum_{\mathbf{q}} \sum_{\alpha} v_{\alpha}(\mathbf{q}) \langle S({\bf q}) \rangle_T
	e^{i\mathbf{q}\cdot\mathbf{r}}
\label{eq-va}
\end{eqnarray}
where ${\bf q}$ is the reciprocal lattice vector. 
$v_{\alpha}({\bf q})$ is the Fourier
transform of the screened atomistic potential at zero temperature, 
which takes the form of, \cite{Williamson}
\begin{equation}\label{eq:vq}
v_{\alpha}(q) =  \frac{\alpha _0(q^2 -\alpha_1)}{\alpha_2 e^{\alpha_3 q^2} -1}
\end{equation}
where $q=|{\bf q}|$, and
the parameters $\alpha_0$, $\alpha_1$, $\alpha_2$, $\alpha_3$ are fitted
to the bulk properties of InAs and GaAs, including the band gaps, band
offsets, effective masses, etc. $\langle S_{\alpha}(\mathbf{q}) \rangle_T$  is 
the averaged structure factor over all the phonon
configurations at temperature $T$,
\begin{equation}
\langle S_{\alpha}(\mathbf{q}) \rangle_T = \langle \sum_{n}
e^{-i\mathbf{q}\cdot\mathbf{R}_{n\alpha}} \rangle_T \, .
\end{equation}
Assuming ${\bf R}_{n\alpha} = 
{\bf R}_{n\alpha}^0 + {\bf u}_{\alpha}$, where ${\bf u}_{\alpha}$ is the amplitude 
of the phonon mode.
For any ${\bf q}$, we have
\begin{equation}
\langle  e^{-i {\bf q}\cdot {\bf u}_\alpha} \rangle_T = e^{-{\frac{1}{2}}
\langle ({\bf q}\cdot {\bf u}_\alpha)^2\rangle_T}.
\end{equation}
using Wick's theorem. \cite{marder} Therefore, the temperature effects to the
atomic potentials are equivalent to consider a temperature dependent structure
factor,
\begin{equation}
\langle S_{\alpha}(\mathbf{q}) \rangle_T=\sum_{n}e^{-i\mathbf{q}\cdot\mathbf{R}_{n\alpha}^0}
e^{-W_{\alpha}(\mathbf{q},T)}\, ,
\label{eq-tS}
\end{equation}
in Eq. \ref{eq-va}, where $W_{\alpha}(\mathbf{q}, T)$ is the dynamical
Debye-Waller factor for the $\alpha$-th element,
\begin{equation}
2W_{\alpha} ({\bf q}, T) = \langle ({\bf q}\cdot {\bf u}_\alpha)^2\rangle_T \,.
\end{equation}
For simplicity, we assume the system to be isotropic, then we have
\begin{equation}
W_{\alpha}(\mathbf{q}, T)=\frac{1}{6}|{\bf q}|^2_{}\langle
u^{2}_{\alpha}\rangle \, ,
\end{equation}
where $\langle u^{2}_{\alpha}\rangle$ is the total mean-square displacement
for atom of type $\alpha$ at temperature $T$, including the
contribution from acoustic ($\text{A}$) and optical ($\text{O}$) phonons,
\begin{equation}
\langle u_{\alpha}^2 \rangle
= \langle u_{\alpha}^2 \rangle_{\text{A}} + \langle u_{\alpha}^2 \rangle_{\text{O}}.
\end{equation}
We use Debye model for acoustic phonons,
\begin{eqnarray}
\langle u^{2} \rangle_{\text{A}}
&& = \int^{\omega_{D}}_{0}g(\omega)\frac{\hbar}{NM\omega}
\left(\frac{1}{e^{\hbar\omega /k_{B}T}-1}+\frac{1}{2}\right)d\omega \nonumber \\
%&& =\frac{9\hbar}{m\omega_D}\left(\frac{T}{\theta_{D}}\right)^{2}
%  \left[\int^{\frac{\theta_{D}}{T}}_{0}\frac{x}{e^{x}-1}dx
%  + \frac{1}{4}\left(\frac{\theta_{D}}{T}\right)^{2}\right] \nonumber \\
&& = \alpha_4\left[\left(\frac{T}{\theta_{D}}\right)^{2}
\int^{\frac{\theta_D}{T}}_{0} \frac{x}{e^{x}-1}dx+\frac{1}{4}\right]\, ,
\label{eq-dw-A}
\end{eqnarray}
where $\alpha_4 = 9\hbar/M\omega_D$, and $\theta_D=\hbar\omega_D$ is the Debye
temperature.  The second term in the bracket is from zero point quantum
fluctuation, which can be absorbed to the zero-temperature pseudopotentials.  At
high temperature, the acoustic phonon displacement $\langle u^{2} \rangle_A
\propto T$, whereas in the low-temperature limit, $\langle u^{2} \rangle_A
\propto T^2$.  For the optical phonons, the average displacement can be written as
\begin{equation}
\langle u^{2} \rangle_{\text{O}} = \frac{\alpha_5}{e^{\hbar\omega_{\text{O}}/K_BT}-1}\, ,
\label{eq-dw-O}
\end{equation}
where $\omega_\text{O}$ is the frequency of the optical phonon. We neglect the
dispersion of optical phonons here.  Generally, the contribution of the optical
phonon could be significant only at high temperature. \cite{fan} Although
$\alpha_4$ and $\alpha_5$ can be calculated directly using the parameters of
bulk materials, the approximations made during the derivations of Eq.
(\ref{eq-dw-A}) and (\ref{eq-dw-O}) can introduce some errors to the energy gap
of semiconductors at finite temperature. To overcome this problem, we treat
$\alpha_4$ and $\alpha_5$ as fitting parameters, which are fitted to the
temperature dependent band structures in combination with the zero-temperature
empirical pseudopotentials. Therefore, the dynamical Debye-Waller factors might
be different from the real physical Debye-Waller factors of the system.

\begin{table}[htbp]
\setlength{\tabcolsep}{1.0em}
\centering
\caption{Band parameters obtained from the empirical pseudopotential
calculations compared with the experimental values. \cite{VM, Otfried:2003}
Energies are in unit of eV. $m^{\ast}_{e}$, $m^{\ast}_{hh}[100]$,
$m^{\ast}_{hh}[111]$ and $m^{\ast}_{lh}[100]$ are effective masses of electron
and holes at $\Gamma$ point. $a_{\mathrm{gap}}$ and $a_{\Gamma_{15v}}$
denote the deformation potential of the band gap and the $\Gamma_{15v}$ point.
$b$ is the valance band biaxial deformation
potential.  $\Delta_{0}$ and $\Delta_{1}$ are the spin-orbit splittings at the
$\Gamma_{15v}$ and $L_{1v}$ points respectivity.}
\label{table-epm}
\begin{tabular}{lrrrrr}
\hline\hline
& \multicolumn{2}{c}{GaAs} && \multicolumn{2}{c}{InAs}\\
\cline{2-3}\cline{5-6}
Property             & EPM    & Expt.  && EPM    & Expt.\\
\hline
$E_{\mathrm{gap}}$   &  1.528 &  1.52  &&  0.423 &  0.42\\
$E_{X_{5v}}$         & -2.763 & -2.96  && -2.365 & -2.40\\
$E_{X_{1c}}$         &  1.937 &  1.98  &&  2.069 &  2.34\\
$E_{X_{3c}}$         &  2.232 &  2.50  &&  2.514 &  2.54\\
$E_{L_{3v}}$         & -1.041 & -1.30  && -0.872 & -1.26\\
$E_{L_{1c}}$         &  2.232 &  1.81  &&  1.568 &  1.71\\
$m^{\ast}_{e}$       &  0.067 &  0.067 &&  0.023 &  0.023\\
$m^{\ast}_{hh}[100]$ &  0.316 &  0.40  &&  0.371 &  0.35\\
$m^{\ast}_{hh}[111]$ &  0.825 &  0.57  &&  0.986 &  0.85\\
$m^{\ast}_{lh}[100]$ &  0.092 &  0.082 &&  0.029 &  0.026\\
$a_{\mathrm{gap}}$   & -7.879 & -8.33  && -6.804 & -5.7\\
$a_{\Gamma_{15v}}$   & -1.110 & -1.0   && -0.829 & -1.0\\
$b$                  & -1.567 & -1.7   && -1.631 & -1.7\\
$\Delta_{0}$         &  0.362 &  0.34  &&  0.384 &  0.39\\
$\Delta_{1}$         &  0.201 &  0.22  &&  0.286 &  0.27\\
\hline\hline
\end{tabular}
\end{table}

\begin{table}
% Underfull \hbox
\setlength{\tabcolsep}{0.4em}
\caption{Fitted pseudopotential parameters for InAs/GaAs in Eq.(\ref{eq:vr})
and Eq.(\ref{eq:vq}). A plane-wave cutoff of 5 Ryd is used.} 
\label{table-alpha}
\begin{tabular}{lrrrr}
\hline \hline
parameters      & Ga       &  As (GaAs)   & In  &  As (InAs) \\
\hline
$\alpha_0$      &476845.70 &11.9753 &771.3695 &26.8882 \\
$\alpha_1$      &1.9102    &3.0181  &1.6443   &2.9716  \\
$\alpha_2$      &22909.50  &1.1098  &18.1342  &1.2437  \\
$\alpha_3$      &0.1900    &0.2453  &0.3940   &0.4276  \\
$\gamma_{\alpha}$&2.5215   &0.0     &2.1531   &0.0     \\
$\alpha_{so}$   &0.1035    &0.0976  &0.5973   &0.0976  \\
\hline \hline
\end{tabular}
\end{table}

\begin{table}[htbp]
\setlength{\tabcolsep}{0.45em}
\centering
\caption{Debye temperatures,\cite{Otfried:2003} the optical phonon energies
and the fitted $\alpha_4$, $\alpha_5$ parameters for GaAs and InAs.}
\begin{tabular}{lllllll} \hline\hline
	& & & \multicolumn{2}{c}{cation} & \multicolumn{2}{c}{anion}\\
\cline{4-7}
Bulk & $\theta_{\text{D}}$(K)  & $\omega_{\text{O}}$(meV)
& $\alpha_4$ & $\alpha_5$
& $\alpha_4$ & $\alpha_5$ \\
\hline
GaAs & 344 & 35.36 &0.3024 &0.0786 &0.1530 &0.0024 \\
InAs & 247 & 29.6  &0.1014 &0.0984 &0.0828 &0.0084 \\
\hline\hline
\end{tabular}
\label{table-a45}
\end{table}

To determine the temperature dependent pseudopotential, we first determine the
pseodupotential parameters $\alpha_0$ - $\alpha_3$ and $\gamma$ at zero
temperature by fitting them to the electronic structures of bulk materials,
including the effective mass, and energies of the high symmetry $\Gamma$, $X$
and $L$ points, etc. The target values and fitted values for GaAs and InAs are
compared in Table \ref{table-epm}, which are in good agreement.  The parameters
$\alpha_0$ - $\alpha_3$ and $\gamma$ are presented in Table \ref{table-alpha}.
With these parameters at hand, we then determine the values of $\alpha_4$ and
$\alpha_5$ by fitting them to the temperature dependent energy gap of bulk
materials, which can be well described by the empirical Varshni formula,
\cite{VM, varshni-parameters}
\begin{equation}
\Delta E_g(T) = - \frac{c_1 T^2}{T + c_2} \, ,
\label{eq-varshni}
\end{equation}
where $c_1$ and $c_2$ are the Varshni parameters.
For GaAs, $c_1$ = 0.5405 meV/K and $c_2$= 204 K and
for InAs, $c_1$ = 0.276 meV/K and $c_2$ = 93 K.\cite{VM}  
The fitted parameters for $\alpha_4$ and $\alpha_5$ are summarized in 
Table. \ref{table-a45}.

The single-particle Hamiltonian (\ref{eq-single-H}) can be solved by expanding
the wave functions into a linear 
combination of Bloch bands (LCBB), \cite{wang99b}
\begin{equation}
\psi_i = \sum_{n, {\bf k}, \lambda} c_{n, {\bf k}, \lambda}^i \psi_{n,\epsilon, {\bf k}, \lambda, T},
\label{eq-psii}
\end{equation}
where $\psi_{n,\epsilon, {\bf k}, \lambda, T}$ is the bulk Bloch bands with
orbital $n$ and wave vector ${\bf k}$ close to $\Gamma$ point at finite
temperature $T$, and $\lambda$ = (InAs, GaAs). 
The experimental lattice constants for InAs and GaAs at given temperature, as
shown in Fig. \ref{fig-lattice}, are given
as input to construct the Bloch basis. At each temperature, we relax the
dot+matrix structure using VFF method to get the the atomic position ${\bf
R}_{n,\alpha}(T)$.

Due to the spatial confinement, the carries in the QDs have strong Coulomb
interactions. The many-particle Hamiltonian read as,
\begin{equation}
H = \sum_{i} \epsilon_i \hat{\psi}_{i}^\dagger \hat{\psi}_{i} + \frac{1}{2} \sum_{ijkl} \Gamma_{ij}^{kl} 
\hat{\psi}_{i}^\dagger \hat{\psi}_{j}^\dagger \hat{\psi}_{k} \hat{\psi}_{l},
\label{eq-Hmp}
\end{equation}
where $\hat{\psi}_{i}=c_i \psi_i({\bf r})$ is the field operator 
with corresponding single particle energy $\epsilon_i$. 
$\Gamma_{ij}^{kl}$ are the Coulomb integrals,
\begin{equation}
\Gamma_{ij}^{kl} = \int\int d{\bf r} d{\bf r'} \frac{ \psi_i^*({\bf r}) \psi_{j}^*({\bf r'})
\psi_k({\bf r'}) \psi_{l}({\bf r})} {\epsilon({\bf r-r'}) |{\bf r-r'}|}.
\end{equation}
Here, $\epsilon({\bf r-r'})$ is the screened dielectric function.
\cite{Franceschetti99} The many-particle Hamiltonian is solved using a
configuration interaction method, \cite{franceschetti00} where the
many-particle wave functions 
are expanded on the Slater determinants constructed from the confined
electron and hole levels. This method has been
successfully applied to studying the electronic and optical properties of
InAs/GaAs QDs and the obtained
results are in very good agreement with the experimental observations.
\cite{Ding, Ediger07a, Ediger07b, Williamson}
%For more details, see Ref. \onlinecite{bester-review}. 

\section{TDEPM For Bulk Materials}
\label{sec-bulk}

\begin{figure}[htbp]
\centering
\includegraphics[width=2in]{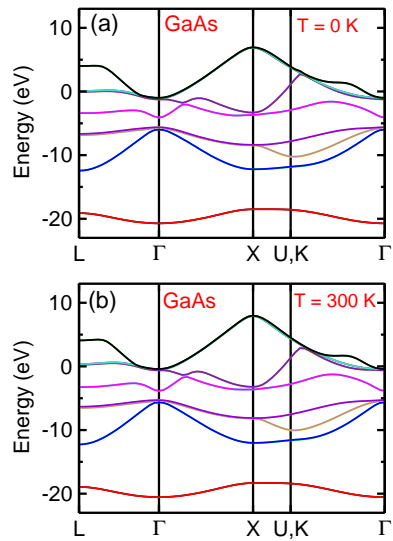}
\caption{(color online) The band structures of GaAs at T = 0 (a) and T = 300 K (b) calculated
by the TDEPM.} \label{fig-GaAs-Band}
\end{figure}

We first test our method for bulk materials.
Figure \ref{fig-GaAs-Band} depicts typical band structures of GaAs at T = 0 and
T = 300 K.  The overall band structures are quite similar to those at zero
temperature even at rather high temperature (300 K).  However, the energies of
high symmetry $k$-points $\Gamma$, $X$, and $L$ have different response to the
temperature.  In principle, all the energies of these $k$-points should be taken
as the target values to determine the values of $\alpha_4$ and $\alpha_5$.
Unfortunately, the experimental data of the energies of these $k$-points at
finite temperature are not available, therefore, tentatively, we fit the
potentials only to the temperature dependent energy gaps at $\Gamma$ point.  The
potentials can be improved by fitting to the energies of more $k$-points in the
future.

\begin{figure}[htbp]
\centering
\includegraphics[width=2.5in]{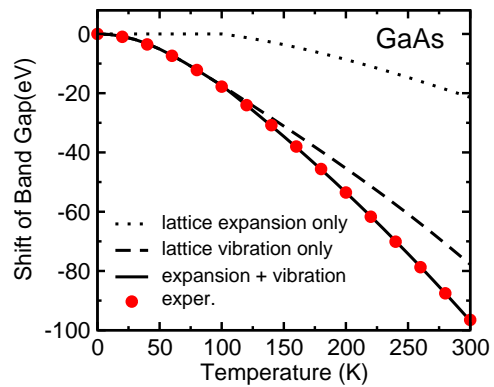}
\caption{(color online) The red shift of band gap as a function of temperature for GaAs. The filled
circles represent the experimental data. The
dashed line is the result calculated from TDEPM with only lattice vibration 
whereas the dotted line is the result with only lattice expansion. 
The solid line is from TDEPM taking account of both lattice vibration and lattice expansion.} 
\label{fig-DeltaEg}
\end{figure}

We present in Fig. \ref{fig-DeltaEg} the change of energy gap $\Delta E_g$ as a
function of temperature for GaAs.  The filled circles are the results taken from
experiments, \cite{VM} whereas the solid line is calculated from TDEPM.  The
fitting error is less than 0.2 meV in the whole temperature range.  The dashed
line is the change of energy gap of GaAs taken account only lattice vibration,
whereas the dotted line is the result with only lattice expansion.  As we see,
to accurately describe the red shift of the band gap with respect to
temperature, both the lattice expansion and lattice vibration have to be taken
into account in the theory. 
These results agree with the
results obtained from temperature dependent tight-binding method by Pour
et al. \cite{abdollahi11} 
Similar features are also found for InAs using the
parameters given in Table \ref{table-a45}.

\begin{figure}[htbp]
\centering
\includegraphics[width=2.5in]{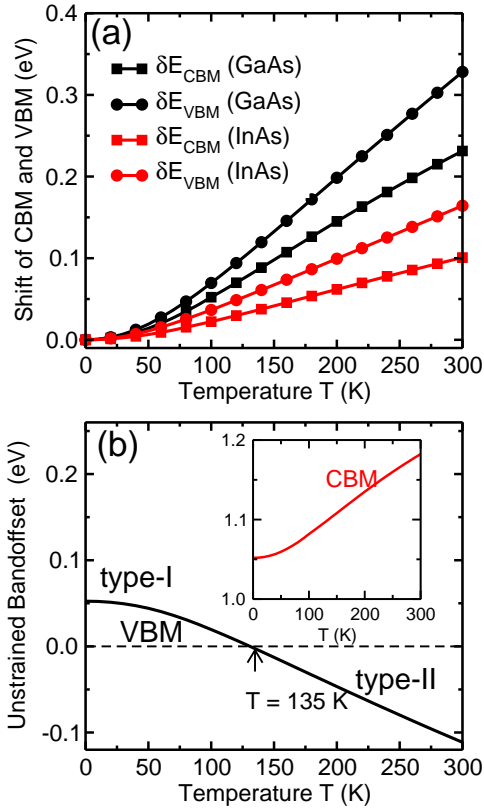}
\caption{(color online) (a) Changes of CBM and VBM as functions of temperature for InAs and
GaAs bulk materials. (b) The valence band offset and the conduction band
offset (inset) between InAs and GaAs as functions of temperature.}
\label{fig-bulkboff}
\end{figure}

The conduction band offset ($\Delta E_c$) and valance band offset ($\Delta E_v$)
between InAs and GaAs, defined as
\begin{eqnarray}
\Delta E_c && =  E_{\Gamma}^{\text{CBM}}(\text{GaAs})  - E_{\Gamma}^{\text{CBM}}(\text{InAs}), \nonumber \\
\Delta E_v && =  E_{\Gamma}^{\text{VBM}}(\text{InAs})  - E_{\Gamma}^{\text{VBM}}(\text{GaAs}),
\label{eq-vbo}
\end{eqnarray}
are very important to the electronic structures of the InAs/GaAs
heterostructures, because they are important for the confinement of electron and
hole in InAs/GaAs QDs.  We show the change of CBM and VBM of InAs and GaAs with
temperature in Fig. \ref{fig-bulkboff} (a), and the temperature dependent band
offsets between InAs and GaAs in  Fig. \ref{fig-bulkboff} (b). We find:

(i) The CBM and VBM of GaAs change much larger than their counterparts in InAs.

(ii) The CBM generally changes much larger than VBM. At high temperature ($T >$
100 K), CBM and VBM change approximately linearly with respect to
temperature. The changes of the band offsets with respect to the temperature are
presented in Fig. \ref{fig-bulkboff} (b) for the VBM and in the inset for CBM.
For electron, the temperature effect generally enhances the band offset, whereas
for hole, the band offset decreases with the increasing of the temperature.
Interestingly, there is a  type-I to type-II transition at $T$ = 135 K for
unstrained InAs/GaAs.  However, becauase of the strain effects, the holes
are still localized in InAs/GaAs QDs (see Sec. \ref{sec-qds}). In what
follows we will show that the change of band offsets at finite temperature
will greatly modify the electronic and optical properties of QDs.

\section{TDEPM For InAs/GaAs QDs}
\label{sec-qds}

\begin{table}
% Underfull \hbox
\setlength{\tabcolsep}{1.2em}
\centering
\caption{The alloy composition and size of the lens shaped In$_x$Ga$_{1-x}$As/GaAs QDs used in this work.}
\begin{tabular}{cccc} \hline\hline
\#     & $x$ & Base (nm) & Height (nm) \\       \hline
QD-A   & 1.0    & 25 & 3.5 \\
QD-B   & 0.6  & 25 & 3.5 \\
QD-C   & 0.7  & 25 & 3.5 \\
QD-D   & 0.8  & 25 & 3.5 \\
QD-E   & 1.0  & 25 & 5.5 \\
QD-F   & 0.6  & 25 & 5.5 \\
QD-G   & 0.7  & 25 & 5.5 \\
QD-H   & 0.8  & 25 & 5.5 \\
\hline\hline
\end{tabular}
\label{table-QDs}
\end{table}

In Sec. \ref{sec-bulk}, we study the temperature dependent electronic structures
of bulk InAs and GaAs. The band offsets between InAs and GaAs are greatly
modified due to the temperature effects and there is a type-I to type-II
transition in the unstrained InAs/GaAs heterostructure. The change of band offsets
will significantly change the corresponding electronic and optical properties in
QDs. However, the temperature dependent properties are more complicated in QDs
because of the strain effects. In this section, we investigate the temperature
dependent electronic and optical properties of InAs/GaAs QDS using TDEPM.  We
study QDs with different sizes and alloy compositions. The alloy
compositions of selected lens-shaped QDs are presented in Table \ref{table-QDs}. In most of
the cases, we use QD-A to illustrate the main physics. The results of other QDs
will also be presented for comparison.

\subsection{Temperature dependent band offset}
\label{subsec-band offset}

\begin{figure}[htbp]
\centering
\includegraphics[width=2in]{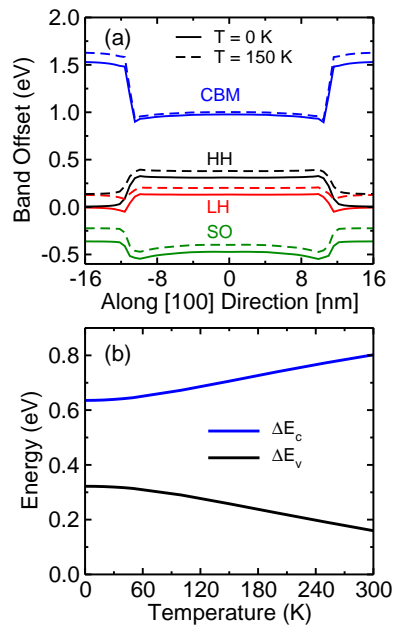}
\caption{(color online) (a) Typical band profiles in strained QDs. HH, LH and SO correspond to
heavy hole, light hole and spin-orbit bands respectively. (b) The band offsets at
the center of InAs/GaAs QDs as functions of temperature calculated
from TDEPM.} \label{fig-boff}
\end{figure}

We first investigate the strained band offsets in InAs/GaAs QDs, which is
crucial for the electronic and optical properties of QDs. In previous works,
\cite{he04a, gong08} the Bir-Pikus model is used to obtain the strain modified
band profiles. However, the temperature dependent parameters for the Bir-Pikus
model is generally unavailable.  Therefore, we calculate the band profiles
directly using the TDEPM.  After the lattice relaxations for the dot system, we
construct the 8-atom unit cell according to the local strain, and then calculate
the band structures using the TDEPM.  Typical band offsets in InAs/GaAs dots
along the [100] direction at $T = 150$ K is compared to those of zero
temperature in Fig. \ref{fig-boff} (a).  
We see that the overall profiles of the heave hole (HH)
and light hole (LH) and spin-orbit (SO) bands are still quite similar at the two
temperatures. 
The SO band is lower than the HH and LH by
about 400 meV in the matrix and is greatly enhanced in the dot materials. The
degeneracy of the HH and LH bands is broken 
because of the biaxial strain. \cite{he04a,wei94} 
The strained band offsets of CBM and VBM in the center of 
InAs/GaAs QDs are presented in Fig.
\ref{fig-boff} (b) in the temperature range of 0 - 300 K.  For electrons, the
confinement is enhanced with increasing of the temperature, whereas for holes,
the confinement decreases from 320 to 180 meV.  However, unlike the bulk
materials, even at high temperature, $\Delta E_v$ is always positive, indicating
that the hole is always localized in the QDs. The change of the band offsets
greatly modifies the electronic and optical properties of QDs, as shall be
discussed below. 

\subsection{Temperature dependent single particle levels and wave functions}
\label{subsec-single}

\begin{figure}[htbp]
\centering
\includegraphics[width=2.2in]{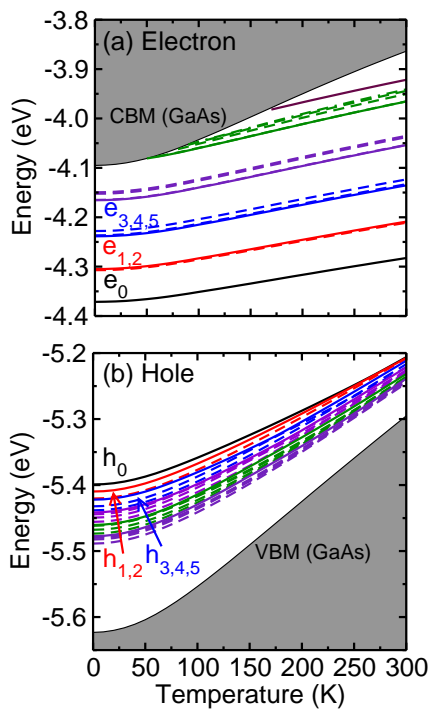}
\caption{(color online) Single-particle energy levels of the confined states as functions of
  temperature in QD-A. The shaded regime at the
top (bottom) panel corresponding to the unconfined states with
energies higher (lower) than the CBM (VBM) of GaAs.}
\label{fig-confined}
\end{figure}

The evolution of single particle energy levels as functions of temperature is
illustrated in Fig. \ref{fig-confined}.  We show the results for QD-A here. 
Similar features are also found for all other QDs. 
We show all confined electron states, and the highest 20 hole states.  Because of the
enhancement of confinement potential for electrons, more states are confined in
the dots as the increasing of the temperature.  The labels of $e_i$ ($h_i)$
represent the energy levels of electron (hole) in ascending (descending) order.
One can also use angular momentum $s$, $p$, $d$, etc. to label the wave
functions.  For instance, the states $e_0$ and $h_0$ can be labeled as $s$ and
the $e_{1,2}$ ($h_{1,2}$) states are usually labeled by $p_{1,2}$, etc.

\begin{figure}[htbp]
\centering
\includegraphics[width=2in]{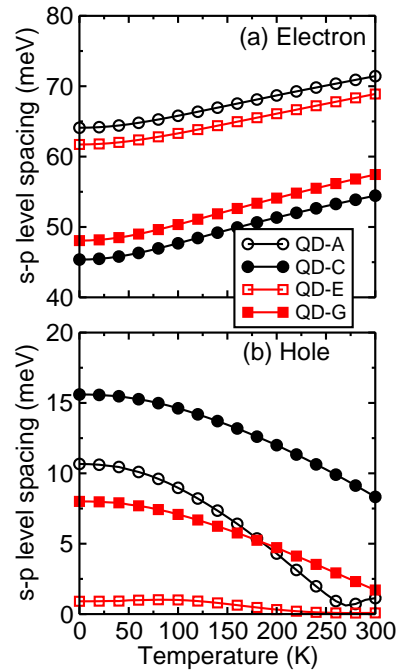}
\caption{(color online) The single-particle $s$-$p$ level energy spacing as functions of
  temperature for electrons and
 holes in the InAs/GaAs QDs.}
\label{fig-sp}
\end{figure}

The $s$-$p$ energy level spacing is shown in Fig. \ref{fig-sp}, which
is defined as,
\begin{equation}
\delta \epsilon_{sp} = |\frac{\epsilon_{p_1} + \epsilon_{p_2}}{2} - \epsilon_{s}|
\end{equation}
for both electrons and holes. Because of the increase of confinement potential,
the electron $s$-$p$ level spacing increases with the increasing of the
temperature. In contrast, for holes, the energy level spacings decreases with the
increasing of the temperature due to the decrease of the hole confinement
potential.  For QD-A, we find that the level spacing of electron increases from
64 to 70 meV when the temperature increases from 0 K to 300 K. For hole, the
energy difference decreases from 11 to about 1 meV. The change of level spacings
in QDs may be measured from PL emission spectra.

\begin{figure}[htbp]
\centering
\includegraphics[width=2in]{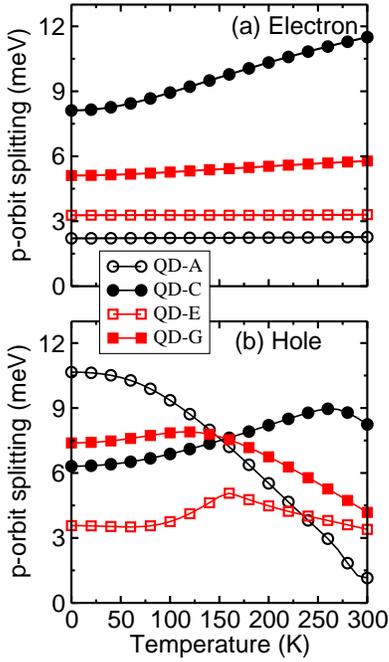}
\caption{(color online) The single-particle $p$-orbit energy splittings as a functions of temperature
for electrons and holes in InAs/GaAs QDs.}
\label{fig-pp}
\end{figure}

The $p$-orbit splitting is another important quantity for the single particle
levels, which is defined as,
\begin{equation*}
\delta \epsilon_{pp} = |\epsilon_{p_1} - \epsilon_{p_2}|.
\end{equation*}
If the QDs have $C_{4v}$ or $T_d$ symmetry, 
%the [110] and [1$\bar{1}$0] directions are
%equivalent, and 
the $p$-orbit splitting is exactly zero.  However, for real QDs,
the highest symmetry is $C_{2v}$, and in alloyed QDs, the symmetry is $C_1$,
the splitting is nonzero.  The results of the $p$-orbit splitting are
presented in Fig. \ref{fig-pp} for different types of QDs. For QD-A and QD-E,
the electron $p$-orbit splittings are almost independent of temperature, whereas
for other two types of QDs the $p$-orbit splittings slightly increase with the
increasing of temperature. 
The results for holes are very different from those for electrons, as shown in
Fig. \ref{fig-pp} (b). Because the hole level spacings are very small and
the anti-crossing between the hole levels may occur when increasing the
temperature, the $p$-level splittings are not monotonic functions of the
temperature. At high temperature, the $p$-level spacing may even exceed
the $s$-$p$ level spacing, which will never happen for electrons.
The $p$-orbit splitting can be measured experimentally via pump-probe
spectroscopy.\cite{zibik04,zibik09}

\begin{figure}[htbp]
\centering
\includegraphics[width=3.2in]{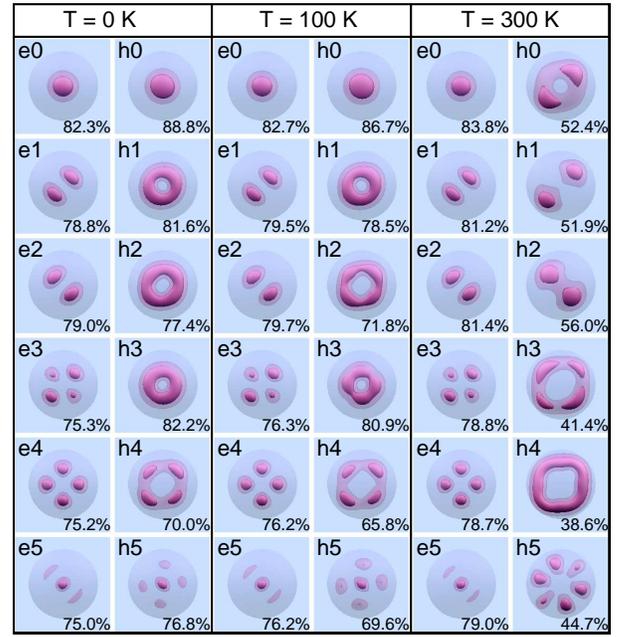}
\caption{(color online) The squared envelope wave functions of the confined electron
and hole states at $T$ = 0 K, 100 K and 300 K. The number on the lower right corner
of each small panel represents the percentage of the state densities
confined in the QDs.}
\label{fig-wf}
\end{figure}

The change of confinement potentials may also change the shapes of the wave
functions.  We present the squared envelope wave functions of electrons and holes at $T$ =
0, 100 and 300 K in Fig. \ref{fig-wf} for QD-A.  The number on the lower
right corner of each small panel represents the percentage of the density 
confined in the QDs.  For $e_{0,1,2}$, we find that the confined 
densities increase by about 1\% - 2\%, whereas for the higher states, the
confined densities may increase about 4\% - 5\%.  
Although the electron squared wave functions confined in QDs
slightly increase, their overall shapes hardly change. In
contrast, for holes, the confined densities reduce dramatically when the temperature
increases.  For instance, the confined density decreases from 88.8\% to
52.4\% for $h_0$ when the temperature increases from 0 to 300 K.  For $h_5$, the
confined density decreases from 76.8\% to 44.7\%. 
The shapes of the envelope functions for holes 
also change dramatically. For instance, the $h_0$ state
is Gaussian-like at low temperature, but at
high temperature there is a node at the center of the wave function. 
This change of the wave function is due to the enhancement of the interfacial effect in
the QDs, because the confinement is very small for holes at high temperature.
At low temperature only the tall QDs have such interfacial hole
states. \cite{he04a,gong08} 
The change of the wave functions with respect to 
temperature can be measured experimentally using magnetotunneling
spectroscopy. \cite{bester07, vdovin}

\subsection{Temperature dependent PL emission spectrum}
\label{subsec-pl}

\begin{figure}[htbp]
\centering
\includegraphics[width=3in]{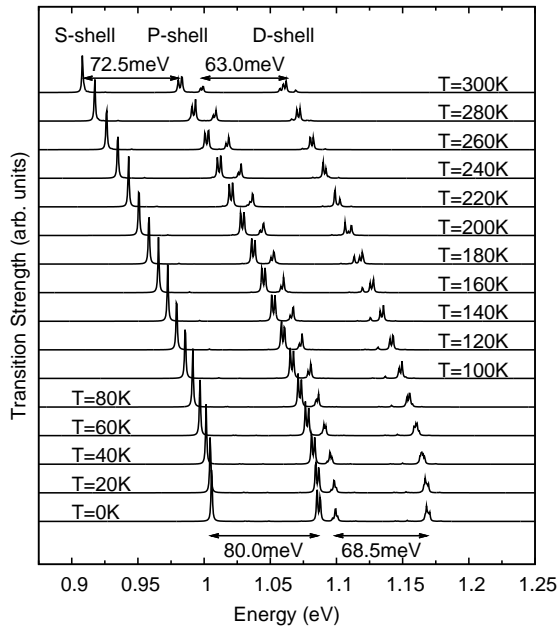}
\caption{Temperature dependent PL emission spectra of QD-A. The emission lines
are broaden by 0.5 meV with a Lorentz function.}
\label{fig-pl}
\end{figure}

In this section, we investigate the temperature dependence of the optical
spectra of InAs/GaAs QDs.  The optical spectra of single InAs/GaAs QD
\cite{ortner, bayer, marzin, yamauchi, rabi, CPT}  have been measured at low
temperature. The highest measuring temperature as far as we know is up to 100 K,
performed by Ortner {\it et al}.\cite{ortner} At higher temperature, the signal to
noise ratio may become very low and the single QD emission is usually hard to
detect. However, for QDs ensemble, the emissions from QDs can be
resolved even at room temperature.  Therefore in this work, we study the optical
spectra of QDs up to 300 K.

Figure \ref{fig-pl} depicts the PL emission spectra of QD-A from 0 K to 300 K.
The energy difference between $S$ and $P$ shell emission can be approximated by
the sum of $s$-$p$ single particle energy level spacing of electron and hole, i.e.
\begin{equation*}
\Delta E_{SP} \approx \delta \epsilon_{sp}^e + \delta \epsilon_{sp}^h\, .
\end{equation*}
Although the change of level spacing in Fig. \ref{fig-sp} can not been
directly measured from PL emission spectra, the sum of them can be measured. The
energy difference between $S$ and $P$ shells for QD-A is 80 meV at zero
temperature and reduces to 72.5 meV at 300 K.

\begin{figure}[htbp]
\centering
\includegraphics[width=3in]{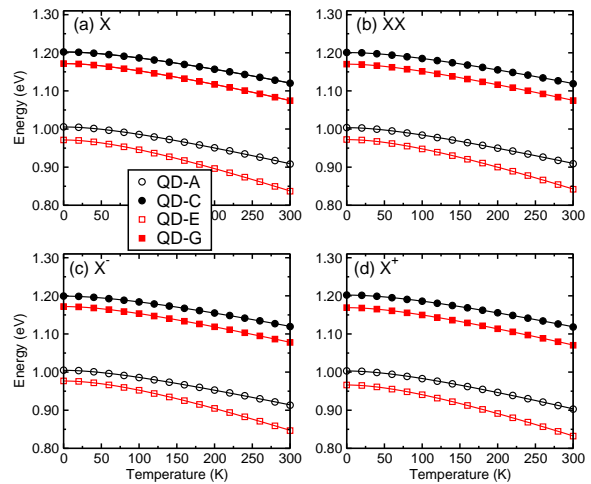}
\caption{(color online) The red shifts of emission lines of (a) X, (b) XX, and 
(c) X$^{-}$ and (d) X$^{+}$
as functions of temperature. The symbols are the results calculated from
TDEPM, whereas the solid lines are the fitted results using Varshni formula.}
\label{fig-varshni4}
\end{figure}

Figure \ref{fig-varshni4} depicts the temperature dependent energies of the
primary exciton, biexiciton and charged excitons for dots A, C, E, G. All the
energies of exciton complexes show red shift as the temperature increases.  For all
dots, we find that the red shifts of the exciton complexes emission lines can be
fitted very well using Varshni formula, with the fitting errors generally less than 1 meV.
The fitting errors are slightly larger than that for the bulk materials ($<$ 0.2
meV), but much smaller than the total red shift of the emission line which is
about 100 meV.  This suggests that the red shift of the emission energies
is proportional to  $T^2$ at low temperature.  The values of the Varshni parameters
are summarized in Table \ref{table-varshni}. These parameters are very
different than those of bulk InAs, GaAs, and also vary from dot
to dot.

\begin{table*}[tbhp]
\caption{Fitted Varshni parameters for different QDs. $E_0$ is the energy at
zero temperature in unit of eV. The Varshni parameters, $c_1$ is in unit of meV/K
and $c_2$ is in unit of K. The geometry and alloy composition of the
QDs are listed in Table \ref{table-QDs}.
}
\begin{tabular}{lllllllllllll} \hline \hline
\# 	  & $E_0(X)$ & $c_1(X)$  & $c_2(X)$  &  $E_0(XX)$ & $c_1(XX)$ &$c_2(XX)$  &$E_0(X^-)$ &$c_1(X^-)$  &$c_2(X^-)$ &$E_0(X^+)$ & $c_1(X^+)$ &$c_2(X^+)$ \\ \hline
 QD-A &   1.0056 &   0.4861 & 151.2457 &   1.0038 &   0.4583 & 137.7505 &   1.0048 &   0.4451 & 139.6235 &   1.0031 &   0.5050 & 155.5669   \\
 QD-B &   1.2600 &   0.4579 & 231.8752 &   1.2592 &   0.4500 & 217.6022 &   1.2568 &   0.4458 & 229.9588 &   1.2610 &   0.4605 & 217.5235   \\
 QD-C &   1.2021 &   0.4419 & 187.7806 &   1.2007 &   0.4364 & 181.1664 &   1.1994 &   0.4266 & 182.4588 &   1.2021 &   0.4493 & 184.1575   \\
 QD-D &   1.1109 &   0.4357 & 157.3439 &   1.1392 &   0.4275 & 151.4595 &   1.1387 &   0.4062 & 141.2489 &   1.1399 &   0.4503 & 161.8518   \\
 QD-E &   0.9714 &   0.7165 & 180.7079 &   0.9725 &   0.7127 & 193.0234 &   0.9768 &   0.7148 & 195.3687 &   0.9663 &   0.7178 & 182.4702   \\
 QD-F &   1.2322 &   0.4957 & 188.3547 &   1.2305 &   0.4854 & 182.7430 &   1.2312 &   0.4836 & 189.6459 &   1.2306 &   0.4994 & 183.6016   \\
 QD-G &   1.1717 &   0.5017 & 165.5773 &   1.1702 &   0.4861 & 159.2807 &   1.1720 &   0.4800 & 159.7797 &   1.1691 &   0.5107 & 166.9354   \\
 QD-H &   1.1085 &   0.5393 & 164.2455 &   1.1077 &   0.5151 & 155.4030 &   1.1102 &   0.5059 & 153.0005 &   1.1052 &   0.5424 & 161.4847   \\ \hline\hline
\end{tabular}
\label{table-varshni}
\end{table*}

\begin{figure}[htbp]
\centering
\includegraphics[width=2.7in]{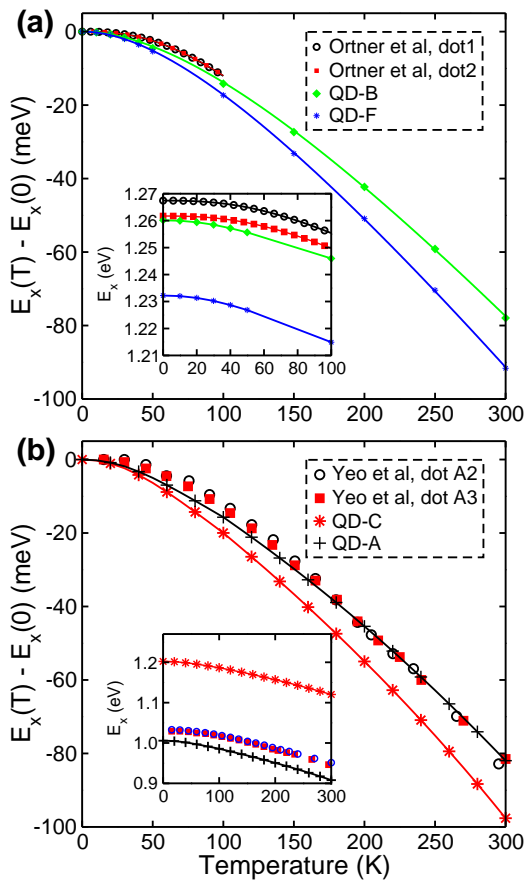} 
\caption{(color online) (a) Comparison of the red shifts of the exciton emission energies
of QD-B and QD-F to the experimental data of single QDs (dot1, dot2) 
in Ref. \onlinecite{ortner}. 
(b) Comparison of the red shifts of the exciton emission energies 
of QD-A and QD-C to the
experimental data of QDs ensemble ($A_2$, $A_3$) in Ref. \onlinecite{Yeo11}.
In both figures, the insets depict the exciton energies as functions of
temperature.} \label{fig-exp}
\end{figure}

The temperature dependent optical spectra has been studied
experimentally by several groups. \cite{ortner, Yeo11} For example, red shift of
exciton emission line in single QDs have been investigated by Ortner {\it et
al}\cite{ortner} from $T$ = 0 K to $T$= 100 K.  In Fig. \ref{fig-exp} (a), we
compare our theoretical results (QD-B, QD-F) with the available experimental
data\cite{ortner} for the In$_{0.6}$Ga$_{0.4}$As/GaAs QDs. The red shift of the
emission line of exciton agree well with QD-B (less than 1 meV).  In the inset
of Fig. \ref{fig-exp}, we present the exciton energies as a functions of
temperature, which also show excellent agreement. The red shift
of exciton energy in Ref. \onlinecite{ortner} can be well described by
Varshni formula using $c_1 = 0.4419$ meV/K and $c_2 = 221.77$ K, with error
less than 1 meV.  This is also in a good agreement with the theoretical values
for QD-B given in Table \ref{table-varshni}.

In Fig. \ref{fig-exp} (b), we compare the theoretical results of QD-A and QD-C
with the experimental results for QDs ensemble measured by Yeo et al.
\cite{Yeo11} 
where the peak energy of $S$-shell is chosen as the  exciton emission lines at each
temperature. For QDs ensemble, the $S$-shell can be well resolved even at
room temperature. We see that the red shifts of QDs ensemble $A_2$ and $A_3$ agree well
with the theoretical prediction of QD-A. Moreover, the exciton energy of $A_2$
and $A_3$ at zero temperature is around 1.02 - 1.03 eV, also agree well with the
exciton energy of QD-A listed in Table \ref{table-varshni}.

We note that some experiments\cite{cardona} suggest that the redshift of the
exciton energies proportional to $T^4$, \cite{passler99,fernandez97} 
instead of $T^2$ as suggested by TDEPM at
very low temperature ($T < 10$ K). The discrepancy may come from two reasons.
First, it is because we do not have high accurate temperature dependent band
gaps to fit at very low temperature at present stage.  It may also partly comes
from the approximations we made in deriving the temperature dependent
pseudopotentials. Nevertheless, in this temperature range, the change of exciton
energies is very small, and the difference between the experimental values and
the theory is very subtle.    

We also calculate the FSS and the polarization of the mono-exciton at finite
temperatures. We find that the FSS and the polarization are generally
insensitive to the temperature.  For example, the change of FSS is usually less
than 1 $\mu$eV, and the change of polarization is less than 5 degree when
increasing the temperature from 0 to 100 K. This result suggests that the FSS can
not been tuned using temperature effect.

\section{Summary and Conclusions}
\label{sec-summary}

We develop a temperature dependent empirical pseudopotential theory,
and apply it to study
the electronic and optical properties of self-assembled InAs/GaAs quantum dots (QDs) 
at finite temperature. The theory takes the effects of 
both lattice expansion and lattice vibration into account. 
The pseudopotentials correctly reproduce the temperature dependent band gap of
bulk III-V semiconductors such as InAs, and GaAs, etc.
We find that for the unstrained InAs/GaAs heterostructure,
the conduction band offset increases whereas
the valence band offsets decreases with the increasing of the temperature, and
there is a type-I to type-II transition at approximately 135 K.
Yet, for InAs/GaAs QDs, the holes are still localized
in the QDs even at room temperature because the large lattice mismatch between InAs and GaAs
greatly enhances the valence band offset.
The single particle energy levels in the QDs show strong temperature
dependence due to the change of confinement potentials. As a consequence, more
electron states are confined at higher temperature.
Because of the changes of the band offsets, the electron wave functions confined in QDs
increase by about 1 - 5\%, whereas the hole wave functions decrease by about
30 - 40\% when the temperature increase from 0 to 300 K.
The calculated recombination energies of exciton, biexciton and charged
excitons show red shift with the increasing of the temperature, which are in
excellent agreement with available experimental data. We expect the theory can
facilitate the future device applications of QDs.

{\it Acknowledgments.} - LH acknowledges the support from the Chinese National
Fundamental Research Program 2011CB921200,
National Natural Science Funds for Distinguished Young Scholars
and the Fundamental Research Funds for the Central Universities
No. WK2470000006.

%\bibliographystyle{apsrev}
%\bibliography{ref}

\end{document}